**PAPER • OPEN ACCESS**

# Carbon burning at stellar energies

To cite this article: S. Courtin *et al* 2023 *J. Phys.: Conf. Ser.* **2586** 012114

View the article online for updates and enhancements.

## You may also like

- A SYSTEMATIC STUDY OF CARBON–OXYGEN WHITE DWARF MERGERS: MASS COMBINATIONS FOR TYPE Ia SUPERNOVAE
  Yushi Sato, Naohito Nakasato, Ataru Tanikawa et al.

- Electron Capture Supernovae from Close Binary Systems
  Arend J. T. Poelarends, Scott Wurtz, James Tarka et al.

- THE CRITICAL MASS RATIO OF DOUBLE WHITE DWARF BINARIES FOR VIOLENT MERGER-INDUCED TYPE IA SUPERNOVA EXPLOSIONS
  Yushi Sato, Naohito Nakasato, Ataru Tanikawa et al.

# Carbon burning at stellar energies

**S. Courtin, M. Heine, E. Monpribat, J. Nippert**

Université de Strasbourg, CNRS, IPHC UMR 7178, F-67000 Strasbourg, France

Sandrine.Courtin@iphc.cnrs.fr

**Abstract**. Fusion reactions with light nuclei play an essential role in understanding the energy production, the nucleosynthesis of chemical elements and the evolution of massive stars. The measurement of key fusion reactions at stellar energies is thus of interest, but highly challenging since the associated cross sections are extremely small, of the sub-nanobarn range. Among these reactions, the fusion of carbon nuclei, which drives the stellar carbon burning phase, is deeply connected with essential microscopic features such as the impact of symmetries, the access to quantum states, emerging of resonances or Pauli repulsion effects. These may manifest themselves in exceptional behaviour of the *S*-factor of this particular system and the precision of extrapolations to deep sub-barrier energies is limited. The present contribution discusses recent experimental results of the nuclear astrophysics community on the measurement of the carbon + carbon fusion reaction down to the astrophysics region. The interplay between nuclear structure, nucleosynthesis and stellar evolution is addressed.

## 1. Introduction, Carbon burning

Fusion of two nuclei is one of the principal mechanisms of stellar nucleosynthesis of the elements so that besides the stellar inventory of elements the energy production and hydrostatics conditions depend sensitively on such cross sections. The impact of these fusion reactions depends on their *Q*-value, the mass range that is bridged between seed and target-nuclei and a variety of nuclear structure as well as nuclear-reaction characteristics. In that sense, the fusion of two carbon nuclei towards a magnesium compound, so-called *carbon fusion* is outstanding and unique. These reactions are key in quiescent burning in contracting cores of stars adapting to the gravitational conditions [1] and in more violent events during Type Ia supernovae [2] and likely in superbursts of x-ray binary systems [3]. In the former case, carbon fusion dictates the element flow and energy generation right at the start of advanced burning in massive stars setting the conditions for all subsequent mechanisms. In the latter case, the robust emission characteristics of Type Ia supernovae serve as a cosmological standard for distance measurements.

On the synthesis path towards heavier nuclei, a variety of processes take place one after another within a reaction network that can evolve complex interwoven patterns. The produced element species themselves can, depending on the released energy, interact with the mother element species so that loops with catalytic effects can establish. The most prominent of these processes is the Carbon-Nitrogen-Oxygen cycle (CNO) [4, 5] that entirely dominates the early evolution of heavier stars. During the subsequent stellar evolution, bottle necks of the nucleosynthesis flow mark another class of unique reactions for stellar evolution, where $^{12}C+\alpha$ and $^{12}C+^{12}C$ are acting during the helium and carbon burning phase, respectively.

The cross sections of both reactions are known to reveal various *resonant structures* that are linked to nuclear structure or nuclear reaction effects [6, 7]. The associated *S*-factor is largely enhanced at the resonance energy, which massively increases the stellar reaction rates with immediate coupling to the stellar hydrodynamics equilibrium and element production. As for the Region of Interest (RoI) of $^{12}C+^{12}C$ for heavy stars (see details below), recent nucleosynthesis simulations are taking into account a previously detected resonance at 2.14 MeV while determining the relative energy range currently covered by direct measurements as well as the effect of extreme α-branching of the resonance decay over proton emission in [8]. The impact of an additional hypothetical resonance at 1.5 MeV was investigated in [9], where the *s*- and *p*-process yields tend to increase significantly by activating the $^{13}C(\alpha, n)^{16}O$ reaction.

Deep sub-Coulomb barrier fusion cross-section of systems with negative *Q*-values must be suppressed for energy conservation reasons and the effect was first detected for heavy ion fusion only twenty years ago [10]. Nowadays, a more complete systematic of *fusion hindrance* towards light systems relevant for stellar burning exists [11], but we are still far from a comprehensive assessment of the phenomenon. The impact of fusion hindrance on cross-section extrapolations into the astrophysics RoI is, however, gigantic and sparked numerous elaborate experimental campaigns meant to approach energies relevant for astrophysics. The effect on stellar evolution was studied applying a generic parametrization from systematic [12, 13] and by interpolating a response function between experimental data [8] of the cross-section excitation function. The reduced reaction rates lead to drastic changes of the temperature and density condition during carbon burning resulting in higher ignition temperatures and the counter-intuitive finding of a significantly shorter carbon burning phase.

## 2. Experimental challenges and techniques

The energies relevant for quiescent burning span from 1.2 MeV to about 1.7 MeV for 8/10 solar mass ($M_\odot$) stars and from 1.7 MeV to 2.7 MeV for 25 $M_\odot$, corresponding to stellar temperatures of 0.5 and 0.9 GK, respectively. These relative energies are in the ultra sub-Coulomb regime of the $^{12}C+^{12}C$ system--the interacting nuclei have to tunnel through the effective Coulomb barrier with exponentially dropping penetration probability--so that with *direct measurements* vanishing cross sections (sub nano-barn) have to be reliably detected. Therefore, lately dedicated low counting-rate experiments were designed for investigating carbon fusion [14, 15, 16] with the aims of:

- reducing randoms applying coincident selection criteria of the reaction channel,
- withstanding high heavy-light ion beam intensity with thin target foils or thick targets,
- reliable entry selection during long data taking periods.

These setups combine the analysis of gamma- and charged-particle detector spectra to identity the $^{12}C(^{12}C, \alpha 1)^{20}Ne$ and $^{12}C(^{12}C, p1)^{23}Na$ reactions, that are the most important ones in reach of direct coincidence approaches for investigating carbon fusion at astrophysics energies, where the cross sections are dropping exponentially in the energy loss interval with the target. Using thin targets, either polygonal chains [17] or suited response functions [18] are utilized to account for the drastic cross section variations. In 'thick target experiments', the beam is stopped within the target volume and the cross sections is extracted from subtracting counting yields during a continuous series of measurements with adapted energy spacing.

*Indirect measurements* are meant to circumvent these experimental difficulties and were performed for carbon burning with a Trojan Horse Method (THM) experiment taking advantage of the larger phase space of the two step process *A(a, b)B*, where *a = xs* has a cluster structure with the spectator *s*, which separates from the actual reactant of interest *x* in a first quasi-free breakup. During the analysis, the calculations depend on potentials within the nuclei and the reaction of interest, *A(x, b)B*, can then be extracted from less challenging measurements (see for example, [19, 20]). The THM relies on some preconditions, *i.e. xs* is well clustered, quasi-free conditions have to be realized, the differential cross section is assumed to have a simple, factorized form, and finally the extracted cross sections have to be normalized to directly measured ones at different energies.

## 3. Recent results

Carbon-burning at stellar energies has known a renewed interest since the last INPC Conference due to several important results, some using the THM and more recently using direct reactions with novel technical developments mentioned above. The latter have allowed to step in the $^{12}$C+$^{12}$C Gamow window with unprecedented accuracy.

### 3.1. Indirect reactions

The $^{12}$C+$^{12}$C reaction has been investigated at the INFN, Laboratori Nazionali del Sud, Catania, Italy, using the $^{12}$C($^{14}$N, $\alpha^{20}$Ne)$^2$H and $^{12}$C($^{14}$N, p$^{23}$Na)$^2$H indirect 3-body processes. THM cross sections which have been deduced show several resonances with spins $0^+$, $2^+$, $1^-$, $3^-$, $5^-$. The corresponding reaction rate was found to be largely enhanced. Impact has been found on the ignition of carbon burning in massive stars as well as superbursts of binary systems [21]. A complementary article was published later on this experiment [22].

As mentioned above, THM cross sections have been normalized to direct data, here in a low-energy window of $E_{cm}$ = 2.50–2.63 MeV.

Recent results of such direct measurements are described in the next paragraph.

### 3.2. Direct measurements

Measuring direct excitation functions is extremely challenging at energies of astrophysical relevance. The use of particle-gamma coincidence technique has allowed to revisit such excitation functions and to provide reliable data down to the astrophysical ROI. In particular in 2020, results were published by University of Notre-Dame in the USA and the STELLA collaboration in Europe [23, 18].

At the University of Notre Dame, the $^{12}$C+$^{12}$C reaction was measured using particle-$\gamma$ coincidence techniques with the SAND silicon detector array placed at the high-intensity 5U Pelletron accelerator. The differential thick-target approach was used with steps of 50 keV. Hindrance of the fusion cross-section is observed in disagreement with results of THM mentioned above.

The STELLA collaboration has developed a mobile measurement station for nuclear fusion reactions of astrophysical interest. The setup is based on a fast-rotating thin targets system for heat dissipation, a high-vacuum ($10^{-8}$ mbar) reaction chamber for minimal carbon build-up, double-sided silicon-strip detectors for efficient particle detection and an array of 36 LaBr3 detectors from the FATIMA (FAst TIMing Array) collaboration for efficient and accurate gamma detection enabling nanosecond timing coincident event selection. The setup is situated at the Andromede facility (IJCLab, Orsay, France) which can deliver high-intensity (10 p$\mu$A) and low-energy carbon beams. The setup is fully described in [15]. It has allowed to measure $^{12}$C+$^{12}$C excitation function from the Coulomb barrier down to 2.1 MeV, over 8 orders of magnitude, with extremely low background due to excellent vacuum, particle-gamma coincidences and nanosecond timing. Results of the first $^{12}$C+$^{12}$C experimental campaign are published in [18] and the corresponding *S*-factors are shown in Figure 1. Reliable excitation functions have been obtained for the $^{12}$C+$^{12}$C fusion reaction over 8 down to cross sections of 130 pb. Three regimes have been explored:

- At moderate sub-barrier energies, the STELLA experimental concept has been validated, comparing the STELLA results to previous measurements.
- At deep sub-barrier energies, hindrance of the fusion cross-section has been observed.
- At further lower energies, entering the $^{12}$C-$^{12}$C Gamow window for 25 M$_\odot$ stars, another regime appears, based on an increase of the *S*-factor.

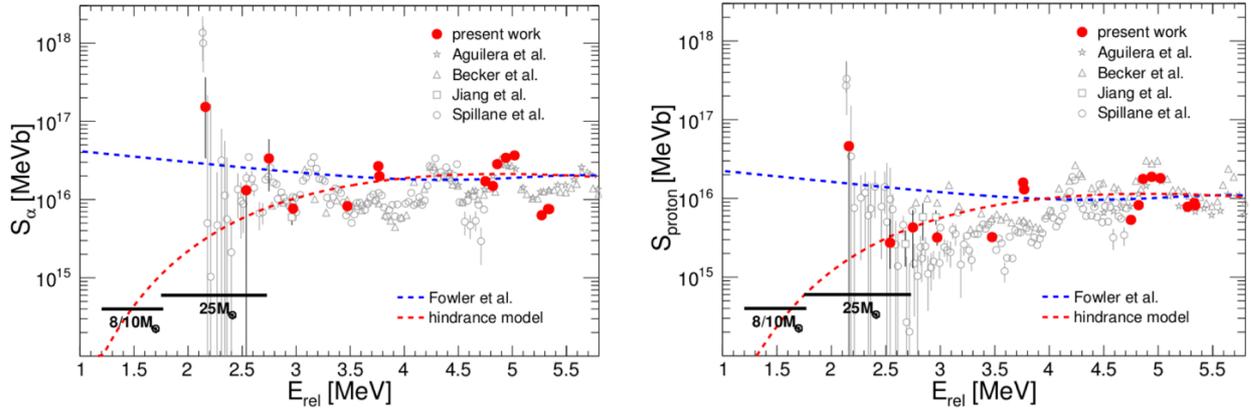

**Figure 1**: S-factor measurements for the $^{12}$C + $^{12}$C fusion reaction as a function of the relative energy $E_{rel}$. See [18] for more details.

The Impact of these new results on stellar evolution and nucleosynthesis is thoroughly investigated in [8]. Description of the measured excitation functions have been made, based on two fusion scenarios. The first one uses the unbiased parameter space of an empirical model of hindrance fusion from [11] (Hin model), and the second one considers the same Hin model with a resonance at the relative energy $E_{rel}$ = 2.14 MeV proposed by [24] (HinRes model). The fitting parameters were *simultaneously adjusted* on both exit channels and their values are consistent with those from [11]. The HinRes model shows a good compatibility with the STELLA measurements. Reactions rates for the carbon fusion have been determined and are shown in Figure 2. The red and green curve represent the Hin and HinRes models respectively. The blue one presents the commonly used CF88 reaction rate [25]. The orange hatched area indicates the STELLA sensitivity zone, and the shaded areas around the curves show the total uncertainties of the reaction rates, based on the experimental uncertainties of the STELLA measured cross sections.

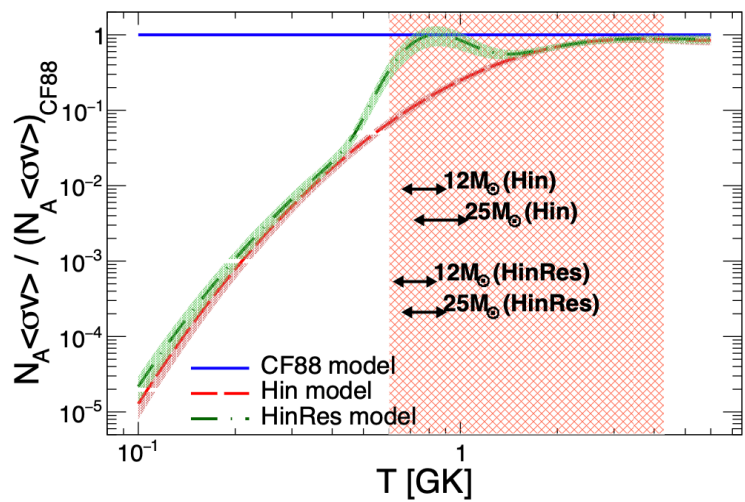

**Figure 2**: Normalized reaction rates to CF88 rates. The orange hatched area indicates the STELLA sensitivity. The black arrows show the regions where carbon fusion occurs for different stellar models, for both Hin and HinRes models. See [8] for more detail.

The reaction rates from Hin and HinRes models are generally lower that the one from CF88. The resonance has an important impact on the rate, where it increases the latter to a level comparable to the one from the CF88 model around $T = 0.85$ GK. Interestingly enough, the STELLA sensitivity corresponds to the temperature range where the reaction rate is determined without extrapolation of the cross section, so that only interpolation between experimental data points is needed.

In order to explore the impact of these reaction rates on stellar evolution, hydrodynamical and nucleosynthesis studies have been performed. First simulations have been made using the Geneva stellar evolution code [26] (GENEC). Two stellar models have been studied: one of 12 $M_\odot$ and the other of 25 $M_\odot$, without rotation and an initial metallicity equal to the solar metallicity. Stellar evolution models have been followed until the end of the carbon burning phase. The main observation is that the fusion temperature for the Hin model is 10% higher than the temperature for the other one, and this reduces the carbon burning lifetime by a factor of two. This can be explained by the fact that the Hin rate is lower than the other ones in the temperature range of the carbon burning. Indeed, to counteract gravitation, a star needs to produce energy, corresponding to a certain reaction rate. Thus, in the case of the Hin model, the star will contract more, become hotter, to reach the relevant reaction rate. This reminds us that a star behaviour is constantly adjusting to gravity. The observed difference in lifetime may have an impact on the neutrino emission, and therefore on the core collapse and the remnant nature. A second observation is that that the carbon burning regions evolve in the same way for both models, but the Hin model shows a convective zone larger than the one for HinRes model. This is due to the higher temperature of the Hin model, and so the presence of a stronger temperature gradient. A study of the nucleosynthesis employing the complete reaction network has been performed with a one-layer code [27]. The abundances obtained at the end of the carbon burning phase show some variations for the sodium, aluminium and phosphorus isotopes, and small variations for heavier elements. These variations may have an impact on the star's evolution [8, 13]. To sum up, the HinRes model is the one compatible with the presented experimental data [8, 18], but in the context of stellar modelling, the comparison with the Hin model demonstrates the impact of extreme branching observed around $E_{rel}$=2.14 MeV [24].

## 4. Overview: symmetry and phase space

On the synthesis path towards heavier nuclei, highly excited compounds are created with various existing processes of energy distribution among their nucleons [28], that can lead to the formation of 'energy concentration' which might be called *clustering* nowadays. The concept was further developed by K. Ikeda associating rotational bands to substructure configurations of alpha conjugate nuclei near decay thresholds [29]. Such phenomena are nowadays calculated from first principles calculations [30] *e.g.* within the framework of anti-symmetrized molecular dynamics (AMD), where detailed information about the resonance energy and spin parity of cluster configurations of $^{24}$Mg were extracted [31]. At iThemba LABS, Cape town, South Africa, a recent study of the $^{24}$Mg($\alpha,\alpha'$)$^{24}$Mg reaction using the K600 Q-2D magnetic spectrometer and the CAKE coincidence array has identified several $0^+$ states in $^{24}$Mg, close to the $^{12}$C-$^{12}$C threshold which predominantly decay to $^{20}$Ne ground state with α emission. In remarkable agreement with results of AMD calculations, these states are discussed to have a dominant $^{12}$C - $^{12}$C cluster structure and are right in the energy region of astrophysical interest of carbon-burning [32]. These low angular momentum states may play a decisive role in the carbon-burning in analogy to the Hoyle state in He-burning and strong impact on stellar modeling may be expected.

In fact, if such states are populated during stellar carbon burning, resonances appear in the fusion excitation function leading to largely enhanced *S*-factors with immediate coupling to the stellar hydrodynamics equilibrium and element production [13].

Finally, hindrance and 'resonant' behavior of the fusion cross-section may be explained using the very same description based on the quantum selectivity of the accessible states for this symmetric identical-boson collision and low density of narrow-width accessible states in the $^{24}$Mg composite system [33].

During the last few years, a step has been made in the Gamow window of Carbon burning with highly reliable direct measurements. In conjunction with indirect methods which may extend to much lower energies, these measurements have paved the way to the accurate exploration of heavy-ion fusion reactions in the energy region of astrophysics interest. This exploration relies on collaborations between nuclear physicists and astrophysicists strongly supported by scientific networks, like IRENA [34] or ChETEC-INFRA [35].

Obviously, new systems may be addressed with possibly new techniques or setups, for example at underground facilities to decrease environmental background. Such projects were addressed during this INPC 2022 conference.

**Acknowledgments**
The authors would like to thank warmly M. Wiescher, W. Tan, A. Tumino, G. Imbriani, P. Adsley and D. Bemmerer for fruitful scientific discussions.
STELLA was funded by the University of Strasbourg IdEX program and CNRS Strasbourg. JN acknowledges support from the Interdisciplinary Thematic Institute QMat, as part of the ITI 2021-2028 program of the University of Strasbourg, CNRS and Inserm, which was supported by IdEx Unistra (ANR 10 IDEX 0002), and by SFRI STRAT'US project (ANR 20 SFRI 0012) and EUR QMAT ANR-17-EURE-0024 under the framework of the French Investments for the Future Program. SC acknowledges the support of the European Union's Horizon 2020 research and innovation programme (ChETEC-INFRA – Project no. 101008324).